\documentclass[FBSedit,FBSmath,ecsub]{FBSsuppl}
\usepackage{amsfonts}
\usepackage{amssymb}
\usepackage{graphics}

\title{Helicity Formalism for NN Scattering without Partial Wave Decomposition}
\author{I. Fachruddin\thanks{\textit{Permanent address:} Jurusan Fisika, FMIPA, 
Universitas Indonesia, Depok 16424, Indonesia} and W. Gl\"ockle}
\institute{Institut f\"ur Theoretische Physik II, Ruhr-Universit\"at Bochum, 
D-44780 Bochum, Germany}

\begin{document}

\maketitle
\begin{abstract}
At intermediate energies it appears advantageous to work without partial wave decomposition and use directly the momentum vectors. We derived a finite set of coupled Lippmann-Schwinger equations using helicities. They can be used for any type of realistic NN forces, which are given in operator form. As an example we have chosen the Bonn OBEPR NN potential.
\end{abstract}

\section{Introduction}

At low energies up to few tenth of MeV very few partial waves are sufficient to describe NN scattering. However, at intermediate (a few hundred MeV) and higher energies many partial waves are needed. Besides in this energy region the scattering amplitude for a high partial wave oscillates strongly in angle. This suggests T-matrix calculations using directly the momentum vectors.

One-boson-exchange NN potentials are derived in momentum space. Instead of using spin operators with a fixed quantisation axis often these potentials are formulated in helicity representation. Moreover, in high energy region taking relativity into account is unavoidable. Thereby, using spin components along a fixed direction is more complicated than using helicities [1].

There are some works [2-4] preceding ours, which address formulations without partial waves and one of them [3] even using helicities. We propose a modified form of equations in comparison to [3]. We present here the formalism for the two nucleon system avoiding partial waves and using helicities.

In Sect. 2 we describe the Lippmann-Schwinger equations as well as the basis states and the physical T-matrix. In Sect. 3 we show the operators to express NN potentials. In Sect. 4 we show the relation between the partial wave T-matrix and the helicity T-matrix. We conclude in Sect. 5.

\section{Basis States, Lippmann-Schwinger Equation and Physical\\ T-Matrix}

To calculate observables we need T-matrix elements. Firstly we define basis states based on which T-matrix elements are calculated by solving the Lippmann-Schwinger equation. The basis states are
\begin{equation}
|{\vec q}; {\hat q}S\Lambda;t \rangle^{\pi a} \equiv \frac {1}{\sqrt{2}}
     (1-\eta_{\pi}(-)^{S+t})|t \rangle |{\vec q};{\hat q}S\Lambda \rangle_{\pi}
\end{equation}
which are antisymmetric choosing \(|{\vec q}; {\hat q}S\Lambda \rangle_{\pi} \equiv \frac {1}{\sqrt{2}} (1+\eta_{\pi}P) |{\vec q}; {\hat q}S\Lambda \rangle\) and \(|{\vec q}; {\hat q}S\Lambda \rangle \equiv |{\vec q} \rangle |{\hat q}S\Lambda \rangle\). Here \({\vec q}\), \(S\), \(\Lambda\), \(P\) and \(\eta_{\pi}\) are relative momenta, total spin, total helicity, parity operator and parity eigenvalue, respectively.

Based on these basis states the T-matrix element is defined as
\begin{equation}
T^{\pi S t}_{\Lambda' \Lambda} ({\vec q'},{\vec q}) \equiv 
     {^{\pi a} \langle} {\vec q'};{\hat q'}S\Lambda';t|T
     |{\vec q};{\hat q}S\Lambda;t\rangle^{\pi a}
\end{equation}
For \({\hat q} = {\hat z}\) one can show \(T^{\pi S t}_{\Lambda' \Lambda} ({\vec q'},{\vec q}) = 
     e^{i \Lambda (\phi' - \phi)} T^{\pi S t}_{\Lambda' \Lambda} (q',q,\theta')\) where \(T^{\pi S t}_{\Lambda' \Lambda} (q',q,\theta')\) obeys a set of coupled Lippmann-Schwinger equations:
\begin{eqnarray}
\lefteqn{T^{\pi S t}_{\Lambda' \Lambda} (q',q,\theta') = 
     \frac {1}{2 \pi} v^{\pi S t, \Lambda}_{\Lambda' \Lambda} 
     (q',q,\theta',0) + \frac {1}{4} \sum_{\Lambda'' = 0}^{S} 
     2^{\Lambda''}}\nonumber\\
& & \times \int_{0}^{\infty} dq'' q''^{2} \int_{-1}^{1} d(\cos \theta'') 
     v^{\pi S t, \Lambda}_{\Lambda' \Lambda''} (q',q'',\theta',\theta'')
     G_{0}(q'') T^{\pi S t}_{\Lambda'' \Lambda} (q'',q,\theta'') 
\end{eqnarray}
with
\begin{equation}
v^{\pi S t, \Lambda}_{\Lambda' \Lambda''} (q',q'',\theta',\theta'') \equiv
     \int_{0}^{2 \pi} d \phi'' e^{-i \Lambda (\phi' - \phi'')}
     V^{\pi S t}_{\Lambda' \Lambda''} ({\vec q'},{\vec q''})
\end{equation}
For \(S = 0\) it is a single equation but for \(S = 1\) it is a set of two coupled equations. We need to solve it only for helicities \(0\) and \(1\). Figure 1 shows a three dimensional plot of the real part of the half-shell T-matrix for S = 1, parity = even, \(\Lambda' = -1\), \(\Lambda = 1\) at \(E_{lab}\) = 100 MeV.
\begin{figure}[hbt]
\centering{\resizebox*{8cm}{6cm}{\rotatebox{270}
     {\includegraphics{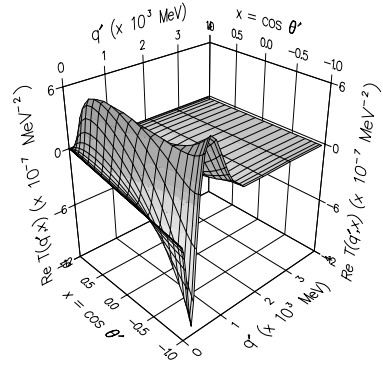}}}}
\caption{The real part of the half-shell T-matrix for S = 1, parity = even, \(\Lambda' = -1\), \(\Lambda = 1\) at \(E_{lab}\) = 100 MeV.}
\end{figure}

The physical T-matrix for spin-magnetic and isospin quantum numbers \(m_{1} m_{2}\), \(\nu_{1} \nu_{2}\) is calculated using the following relation:
\begin{eqnarray}
\lefteqn{{_{a}}\langle \nu_1 \nu_2 m'_1 m'_2 {\vec q'}|T
     |\nu_1 \nu_2 m_1 m_2 {\vec q} \rangle_{a}}\nonumber\\
&& = \frac {1}{4} e^{-i(\Lambda'_{0}\phi'-\Lambda_{0}\phi)}
     \sum_{S \pi t}(1-\eta_{\pi}(-)^{S+t})
     C(\frac {1}{2}\frac {1}{2}t;\nu_{1}\nu_{2})^{2}
     C(\frac {1}{2}\frac {1}{2}S;m'_{1}m'_{2}\Lambda'_{0})\nonumber\\
&& \quad \times C(\frac {1}{2}\frac {1}{2}S;m_{1}m_{2}\Lambda_{0})
     \sum_{\Lambda' \Lambda}d^{S}_{\Lambda'_{0} \Lambda'}(\theta')
     d^{S}_{\Lambda_{0} \Lambda}(\theta)
     T^{\pi S t}_{\Lambda' \Lambda} ({\vec q'},{\vec q})
\end{eqnarray}

\section{Operators for the Potential}

We need potentials given in operator form. For the example of OBEPR there are five types of operators, called \(O\) operators in [3], which cannot be applied directly to our basis states. We take the six \(\Omega\) operators from [3] for which this application is possible, but we redefined \(\Omega_{5}\) so that \(\Omega_{5} = ({\vec S} \cdot {\hat q'})^{2} ({\vec S} \cdot {\hat q})^{2}\). The relation between the \(O\) and the \(\Omega\) operators is then
\begin{eqnarray}
O_{1} & = & \Omega_{1} \qquad \qquad \qquad \qquad
     (\gamma={\hat q'} \cdot {\hat q})\nonumber\\
O_{2} & = & -3\Omega_{1}+2\Omega_{2}\nonumber\\
O_{3} & = & 2{q'^{2}(\Omega_{3}-\frac{1}{2}\Omega_{1})+
     q^{2}(\Omega_{6}-\frac{1}{2}\Omega_{1}) -q'q\Omega_{4}}\nonumber\\
&& +q'q\gamma (2\Omega_{1}-\Omega_{2})+
     q'q\frac{1}{\gamma}(\Omega_{2}-2\Omega_{3}
     -2\Omega_{6}+2\Omega_{5})\nonumber\\
O_{4} & = & \frac{1}{2}q'q(2\Omega_{4}-\gamma \Omega_{2}+\frac{1}{\gamma}
            (\Omega_{2}-2\Omega_{3}-2\Omega_{6}+2\Omega_{5}))\nonumber\\
O_{5} & = & 2{q'^{2}(\Omega_{3}-\frac{1}{2}\Omega_{1})+
     q^{2}(\Omega_{6}-\frac{1}{2}\Omega_{1})+q'q\Omega_{4}}\nonumber\\
&& -q'q\gamma (2\Omega_{1}-\Omega_{2})-
     q'q\frac{1}{\gamma}(\Omega_{2}-2\Omega_{3}
     -2\Omega_{6}+2\Omega_{5})
\end{eqnarray}

\section{The Partial Wave T-Matrix}

From the helicity T-matrix we can calculate the partial wave T-matrix.
\begin{eqnarray} \label{one}
\lefteqn{T^{Sjt}_{l'l}(q) = \frac {4\pi}{(1-\eta_{\pi}(-)^{S+t})
     (1+\eta_{\pi}(-)^{l'})(1+\eta_{\pi}(-)^{l})}\frac{\sqrt{2l'+1}\sqrt{2l+1}}{2j+1}}\nonumber\\
&& \times \sum_{\Lambda' \Lambda}C(l'Sj;0\Lambda') C(lSj;0\Lambda)
     \int_{-1}^{1} d(\cos \theta') d^{j}_{\Lambda \Lambda'}(\theta')
     T^{\pi St}_{\Lambda' \Lambda}(q,q,\theta')
\end{eqnarray}
where \(T^{Sjt}_{l'l}(q) \equiv \langle q(l'S)jmt|T|q(lS)jmt \rangle\). We calculated phase shifts from the T-matrix (\ref{one}) using the Bonn OBEPR potential [5]. They are shown in Table 1 together with those calculated using partial wave decomposition. 
\begin{table}[hbt]
\caption[]{Phase shifts at \(E_{lab}\) = 100 MeV (first line) and \(E_{lab}\) = 325 MeV (second line). \(\delta\) = phase shift from the T-matrix (\ref{one}) and \(\delta_{pw}\) = phase shift from partial wave calculation.}
\centering{\begin{tabular}{|c|r|r|c|r|r|}
\hline
\tabstrut States & \(\delta\) \, \, & \(\delta_{pw}\) \, & States & \(\delta\)
\, \, & \(\delta_{pw}\) \, \\
\hline
\(^{1}S_{0}\) &  27.38 &  27.37 & \(^{3}P_{0}\) &  13.10 &  13.10\\
              &  -7.25 &  -7.26 &               &  -9.43 &  -9.44\\
\(^{1}P_{1}\) & -12.87 & -12.87 & \(^{3}P_{1}\) & -13.72 & -13.72\\
              & -20.09 & -20.10 &               & -28.01 & -28.04\\
\(^{1}D_{2}\) &   3.28 &   3.28 & \(^{3}D_{2}\) &  21.56 &  21.55\\
              &   7.13 &   7.13 &               &  36.25 &  36.20\\
\(^{1}F_{3}\) &  -2.48 &  -2.48 & \(^{3}F_{3}\) &  -1.77 &  -1.77\\
              &  -5.24 &  -5.24 &               &  -4.93 &  -4.93\\
\hline
\end{tabular}}
\end{table}

\section{Summary}

We have formulated a method to carry out scattering process calculations of the two nucleon system without partial wave decomposition in helicity representation. With this method we can reduce the algebraic and numerical work drastically. We need to solve only one set of Lippmann-Schwinger equations instead of solving many for all partial waves. The results of phase shift calculation shows that this method works well.

\begin{acknowledge}
We would like to thank Charlotte Elster from Ohio University, Athens, Ohio for fruitfull discussions and assistance. The first author also would like to thank dem Deutschen Akademischen Austauschdienst for its financial support.
\end{acknowledge}

\end{document}